# Securing Voice-driven Interfaces against Fake (Cloned) Audio Attacks

Hafiz Malik

*Information Systems, Security, and Forensics Lab* (http://issf.umd.umich.edu/)
Department of Electrical and Computer Engineering
University of Michigan – Dearborn
Dearborn, MI 48128, USA
e-mail: hafiz@umich.edu

*Abstract*—Voice cloning technologies have found applications in a variety of areas ranging from personalized speech interfaces to advertisement, robotics, and so on. Existing voice cloning systems are capable of learning speaker characteristics and use trained models to synthesize a person's voice from only a few audio samples. Advances in cloned speech generation technologies are capable of generating perceptually indistinguishable speech from a bona-fide speech. These advances pose new security and privacy threats to voice-driven interfaces and speech-based access control systems. The state-of-the-art speech synthesis technologies use trained or tuned generative models for cloned speech generation. Trained generative models rely on linear operations, learned weights, and excitation source for cloned speech synthesis. These systems leave characteristic artifacts in the synthesized speech. Higher-order spectral analysis is used to capture differentiating attributes between bona-fide and cloned audios. Specifically, quadrature phase coupling (QPC) in the estimated bicoherence, Gaussianity test statistics, and linearity test statistics are used to capture generative model artifacts. Performance of the proposed method is evaluated on cloned audios generated using speaker adaptation- and speaker encoding-based approaches. Experimental results for a dataset consisting of 126 cloned speech and 8 bona-fide speech samples indicate that the proposed method is capable of detecting bona-fide and cloned audios with close to a perfect detection rate.

## I. INTRODUCTION

Artificial human speech synthesis from text, also known as text-to-speech (T2S), is an essential feature in many applications including voice-driven interfaces, humanoid robots, navigation systems, and accessibility for the visually-impaired. Modern T2S systems are based on complex, multi-stage processing pipelines, each of which may rely on hand-engineered features and heuristics. Generative models have been successfully applied to many domains such as image generation [1], speech synthesis [2, 3], and language modeling [4].

Advances in artificial intelligence (AI), speech synthesis, image and video generation technologies pose new security and privacy threats to biometric-based access control systems and voice-driven interfaces. For instance, voice-driven interfaces and services, such as Amazon Alexa [5], Google Home [6], Apple Siri [7], etc., are on the rise. For example, Barclays Wealth has been using passive automatic speaker recognition (ASR) for telephone caller identity verification [8], a verified voiceprint is expected to be used to identify callers. Similarly, the private banking division of Barclays is the first financial services firm to deploy voice biometrics as the primary means to authenticate customers to their call centers [9]. Since then, many voice-biometric-based solutions have been deployed across several financial institutions, including Banco Santander, Royal Bank of Canada, Tangerine Bank, and Manulife, HSBC Bank [10–13]. The existing voice-driven interfaces and voice-biometric-based access control systems are vulnerable to replay and cloned voice attacks, that is, injection of synthesized or recorded voice of an authentic user [14].

This paper presents a framework to secure voice-driven interfaces and services against cloned speech attacks. The proposed method rely on generative model artifacts for cloned speech detection. Specifically, trained generative models rely on linear operation on excitation source and learned weights for cloned speech generation which differs from natural (bona-fide) speech generation process. It is therefore reasonable to assume that cloned speech is expected to exhibits more linearity than a bona-fide speech. Higher-order spectral analysis is used to capture differentiating attributes of bona-fide and cloned audios. Specifically, bicoherence magnitude and phase spectra, Gaussianity test statistics, and linearity test statistics are used to capture generative model artifacts. Performance of the proposed method is evaluated on cloned speech recordings generated using speaker adaptation- and speaker encoding-based approaches discussed in [15]. Experimental results for a dataset of 134 speech samples indicate that the proposed method is capable of detecting bona-fide and cloned audios with close to a perfect detection rate.

### A. Contributions

Contributions of our work are:
1. We have demonstrated that generative models leave characteristics artifacts in the resulting cloned speech recordings.
2. We proposed cloned voice attack model on voice driven-interfaces and automatic speaker recognition (ASR) systems.

3. We proposed a method for cloned audio detection. Higher-order spectral analysis and Gaussianity and linearity statistical tests are used to capture traces of generative models for speech synthesis and human speech.
4. Effectiveness of the proposed method is evaluated on Baidu cloned audio dataset available via [15].

Performance of the proposed is evaluated on cloned speech recordings synthesized using (i) speaker adaptation and (ii) speaker encoding methods discussed in [15]. The performance of the proposed method is also evaluated on voice impersonation using voice morphing via embedding manipulations method discussed in [15].

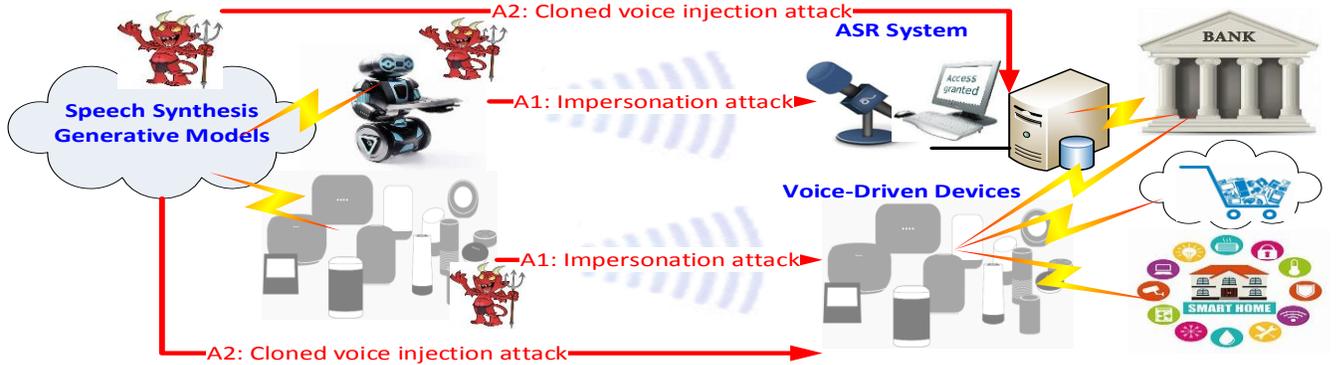

*Figure 1:* Cloned voice attack model for ASR and voice-driven interfaces

## II. CLONED VOICE GENERATION

Recent advances in artificial intelligence (AI) and deep learning has enabled research to clone human speech (e.g., Deep Voice [3, 16, 17]) from few samples [15], images, and video. Cloned speech is an extension traditional text-to-speech (T2S) process. Deep learning based methods [15] train generative models that adopt the same structure of T2S, but differ in replacing all components with neural networks and rely on relatively simpler features. These generative models can be conditioned on text and speaker identity [16] for speech synthesis. For these generative models, text provides linguistic information and controls the content of the generated speech; whereas, speaker identity carries speaker specific characteristics, e.g., pitch, accent, etc. The multi-speaker speech synthesis systems jointly train a generative model and speaker embedding on text, audio and speaker identity [17]. These systems share majority of the model parameters across all speakers and use low-dimensional embedding to encode the speaker-specific information. These methods are capable of generating speech for speakers observed during training phase. Recently, Arık et al [15] proposed few-shot generative modeling of speech conditioned on speaker identity. Their system is capable of voice cloning from few speech samples of an unseen speaker. One of the salient features of the few-shot generative modeling system [15] is that it is capable of voice cloning of a new speaker characteristics from a very limited data, e.g., just few seconds of speech data.

### A. Multi-speaker Generative Modeling for Voice Cloning

Consider a multi-speaker generative model $\varphi(t_{i,j}, s_i; \pi, e_{s_i})$, which takes a text $t_{i,j}$ and a speaker identity, $s_i$, trainable parameters, $\pi$, and trainable speaker embedding corresponding to speaker $i$, $e_{s_i}$. The trainable parameters, $\pi$, and $e_{s_i}$, are optimized by minimizing a loss function, $\mathcal{L}(.)$, expressed as,

$$\min_{\pi,e} E_{\substack{s_i \sim S \\ (t_{i,j}, a_{i,j}) \sim \mathcal{T}_{s_i}}} \{\mathcal{L}(\varphi(t_{i,j}, s_i; \pi, e_{s_i}), a_{i,j})\}$$

where $S$ is a set of speakers, $\mathcal{T}_{s_i}$ is a training set of text-audio pairs for speaker $s_i$, and $a_{i,j}$ is the ground-truth audio for $t_{i,j}$ of speaker $s_i$. The expectation is estimated over text-audio pairs of all training speakers.

Estimates of $\pi$, and e, $\hat{\pi}$, and $\hat{e}$ denote the trained parameters and embeddings. Arık et al [15] proposed two approaches:

*1) Speaker adaptation* which is based on fine-tuning a multi-speaker generative model. Fine-tuning can be applied to either the speaker embedding or the whole model.

*2) Speaker encoding* which directly estimate the speaker embedding from audio samples of an unseen speaker. Such a model does not require any fine-tuning during voice cloning. More details on details of these systems can found in [15] and references therein.

## III. ATTACK MODEL

Cloned voice can be used to attack both ASR-systems and voice-driven interfaces. Shown in Fig. 1 are two possible attack models on both systems.

*1) A1: Impersonation attack where attacker plays cloned speech in using either a smart-speaker or a humanoid robot in front of an ASR system or a voice-driven device.*

*2) A2: Injection attack where attacker directly injects cloned audio into to the target systems an ASR system or a voice-driven device.*

It is important to highlight that injection attack lacks speaker-microphone processing block, therefore, both attacks are expected to introduce two different types of distortions. For example, *injection attack* is expected to more linear than impersonation attack. Moreover, it is also expected to exhibit more linearity when compared with the bona-fide speech. This is mainly because cloned voice generation process is relatively linear than the bona-fide speech generation process which consists of four nonlinear sub-processes: (1) respiration, (2) phonation, (3) resonance, and (4) articulation. Next section outlines a framework for clone voice detection.

### IV. CLONED-VOICE DETECTION USING HIGHER-ORDER SPECTRAL ANALYSIS (HOSA)

We claim that generative models leave characteristic artifacts in resulting cloned speech which can be used for identification. To verify this claim, spectrograms are generated for a ground-truth (bona-fide) speech and corresponding cloned speech generated using speaker adaptation with whole model presented in [15]. Shown in Fig. 2 are the spectrogram plots computed with same set of parameters from bona-fide speech (top) and cloned speech generated using speaker adaptation with whole model approach (bottom). It can be observed from Fig. 2 that cloned speech exhibit vertical lines across time axis artifacts (highlighted using yellow ellipses). Whereas, the spectrogram for bona-fide speech, on the other hand, is smooth across time axis and lacks vertical-line like artifacts. We have observed through extensive experimentation that these artifact are consistent for all cloned speech signals, and these artifacts are consistent irrespective of speech synthesis method. Various approaches can be developed to capture such artifacts. In this paper, we proposed to use higher order spectral analysis (HOSA) based framework to capture these artifacts and used them for cloned speech detection. Motivation behind this choice is two-fold:

(1) Lack of microphone processing block in cloned speech is expected to results in lack of higher order nonlinearity (resp. lack of quadratic phase-coupling (QPC)) in the cloned speech [20].
(2) Cloned speech generation using trained generative model based framework rely on linear operations for speech synthesis. The cloned speech therefore is expected to exhibit higher level of linearity score than a bona-fide speech.

The HOSA-based framework is be used to validate both hypothesis.

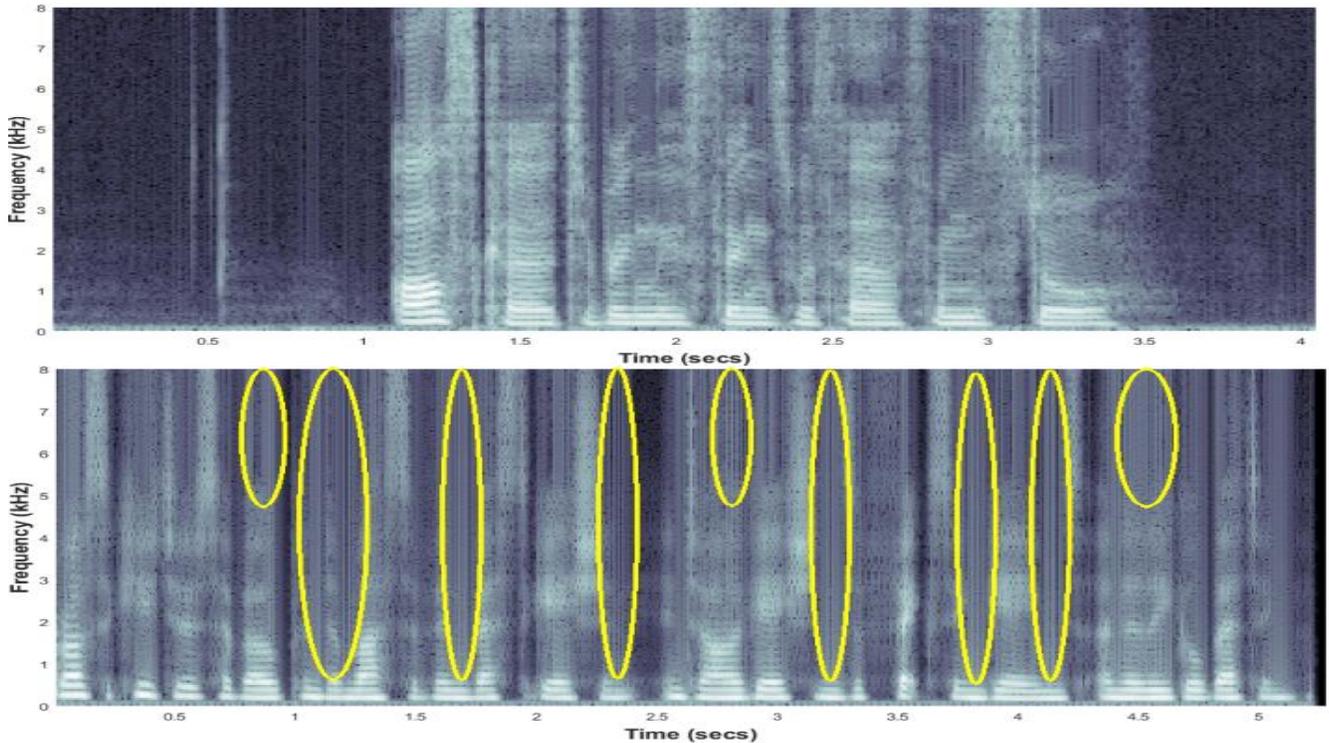

**Figure 2:** Shown is the spectrogram for bona-fide speech (top) and corresponding cloned speech generated using speaker adaptation with whole model (bottom).

#### A. HOSA-based cloned speech detection

Cloned speech lacks microphone processing stage, it is therefore expected to lack microphone/specific distortions such as harmonic–, intermodulation (IM)–, and difference- frequency (DF)–distortions [18,19]. The presence of harmonic components at the output of a nonlinear system with pure tone input is called as *harmonic distortion*. System nonlinearity can cause *IM distortion* in the output when a complex signal (e.g., speech) is applied at the input of a nonlinear system. It causes the output signal to be sums and differences of the input signals fundamental frequencies and their harmonics, that is, $f_1 \pm f_2$, $f_2 \pm 2f_1$, $f_2 \pm 3f_1$, etc. Given a nonlinear system is excited with sum of sinusoids with same magnitudes then system nonlinearity can cause *difference-frequency distortion* at the output, e.g., $2f_2 - f_1$, $2f_1 - f_2$, $3f_1 - 2f_2$, etc.

It has been shown in [18] that microphone is a nonlinear device with a response that can be approximated using following discrete time-invariant Hammerstein series model,

$$y[n] = \sum_k g_1[k]x[n-k] + \sum_\tau g_2[\tau]x[n-\tau]$$

The microphone nonlinearity introduces higher-order correlations at its output. The microphone processing block, therefore, can be modeled using a higher-order nonlinear system. To capture it, HOSA is used. Specifically, higher-order cumulants (resp. bicoherence) [20] is used to capture higher-order correlations. The bicoherence, $(f_1, f_1)$, of a signal $y[n]$ is a normalized version of 2-dimensional Fourier transform of the third-order cumulants, that is,

$$B(f_1, f_2) = \sum_{k_1, k_2 = -\infty}^{\infty} \kappa_y^3(k_1, k_1) e^{-j2\pi(f_1 k_1 + f_2 k_2)}$$
$$= Y(2\pi f_1)Y(2\pi f_2)Y^*(2\pi f_1 + 2\pi f_1)$$

Here, $\kappa_y^3(k_1, k_1)$ denotes third-order cumulant of $y[n]$, and is defined as,

$$\kappa_y^3(k_1, k_1) = E\{y^*[n]y[n+k_1]y[n+k_2]\}$$

Here, E{.} denotes expectation. Sometimes, it is more convenient to use the normalized value of the bispectrum which is also known as bicoherence. This bicoherence is given by the following equation [20],

$$B(f_1, f_2) = \frac{Y(2\pi f_1)Y(2\pi f_2)Y^*(2\pi f_1 + 2\pi f_1)}{|Y(2\pi f_1)Y(2\pi f_2)Y^*(2\pi f_1 + 2\pi f_1)|}$$

It is important to highlight impact of nonlinearity on bicoherence spectrum. Consider a pair of sinusoids with frequencies $f_1$ and $f_2$; the IM distortion will result in a new signal at $f_1 \pm f_2$ whose magnitude is correlated to $f_1$ and $f_2$, which will result in a high magnitude value in the bicoherence magnitude. Moreover, if the input sinusoids have phases, $\theta_1$ and $\theta_2$, then the phase of the nonlinearity induced intermodulation components $f_1 \pm f_2$ are $\theta_1 \pm \theta_2$. It is easy to see that the bicoherence has a zero phase and a bias towards π/2 may also occur due to harmonic auto-correlations. In general, the average bicoherence magnitude would increase as the amount of QPC grows. It is therefore reasonable to assume that cloned speech injection attack is expected to: (i) exhibit relatively smooth magnitude of bicoherence when compared with bona-fide speech, and 2) the phase of bicoherence of cloned speech is expected to be flat and monotonic when compared with bona-fide speech.

To capture traces of a cloned voice injection attack, intermodulation distortion, QPC, Gaussianity test statistics, and linearity statistics can be used. For this paper, QPC, Gaussianity test statistics, and linearity statistics are used. The motivation behind focusing on intermodulation distortion is that it is more dominant in the cloned signal. To verify this claim, we estimated the bicoherence from both the speech and the corresponding cloned recordings. Shown in the top panel of Fig. 3 are the bicoherence phase and magnitude spectra of cloned speech generated using speaker adaptation with whole model and bicoherence phase and magnitude spectra estimated from corresponding ground truth (bona-fide speech).

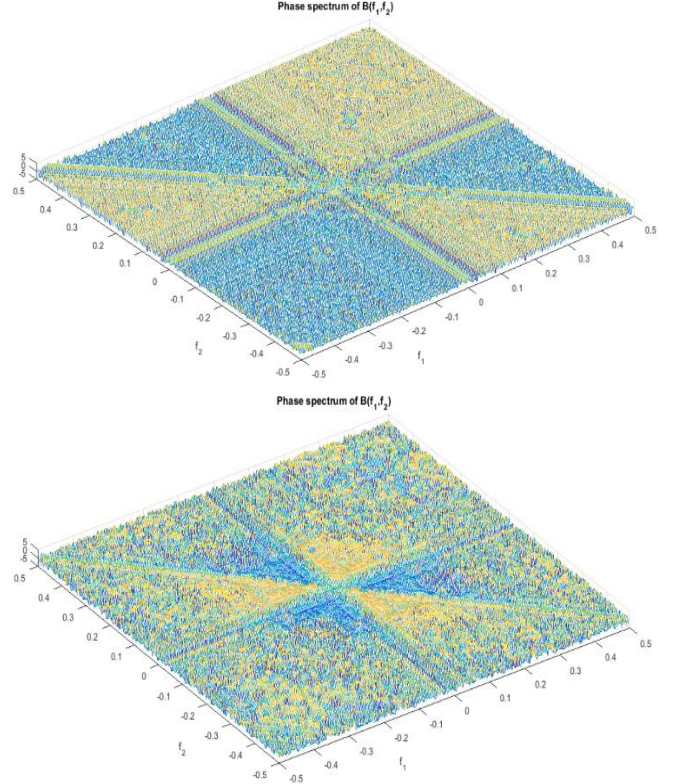

**Figure 3a**: Shown in the top-panel is the phase spectrum of bicoherence estimated from cloned voice using speaker adaptation with whole model and in the bottom-panel is the phase spectrum of bicoherence estimated from corresponding ground truth recoding.

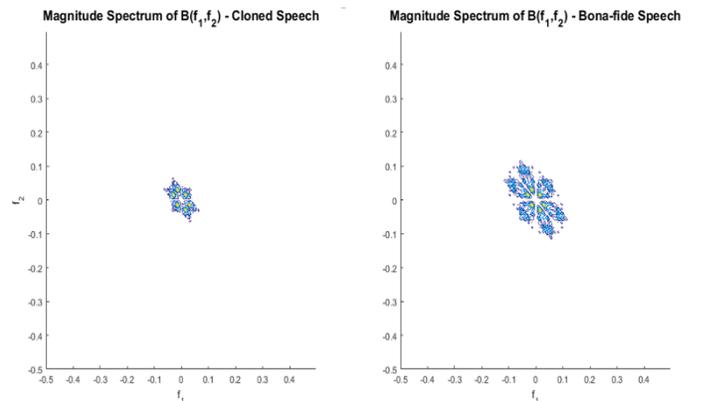

**Figure 3b**: Shown in the left-panel is the magnitude spectrum of bicoherence estimated from cloned voice using speaker adaptation with whole model and in the right-panel is the magnitude spectrum of bicoherence estimated from corresponding ground truth (bond-fide) recoding.

It can be observed from Fig. 3a that there is significant intermodulation distortion spread in the bona-fide speech recording which can be observed from phase spectrum and a non-informative or monotonic bicoherence phase spectrum for cloned speech. It can also be observed from Fig. 3b that there is significant intermodulation distortion spread in the bona-fide speech recording magnitude spectrum which is lacking is the magnitude spectrum of the cloned recording.

## B. Gaussianity test statistics- and linearity test statistics-based detection

Gaussianity and linearity statistics tests can also be used to confirm non-Gaussianity and nonlinearity in a given stationary time series. It is reasonable to assume that bona-fide and cloned speech signals are stationarity sequences. Moreover, bone-fide speech signal is also modeled as a non-Gaussian random sequence. The microphone processing block present in the bona-fide speech is expected to introduce nonlinearity in the resulting sequence.

Let $x(n)$ is non-Gaussian speech sequence and $y(n)$ is linear non-Gaussian sequence of cloned speech. How do we know that $x[n]$ is non-Gaussian and $y[n]$ is non-Gaussina and linear? To achieve this goal, Hinich's non-skewness (also known as Gaussianity) and linearity tests [21] is used.

These tests rely on the fact that if the $3^{rd}$-order cumulants of a stationary process are zero, then its bicoherence is zero, and non-zero bicoherence implies that process is non-Gaussian. Moreover, if that the process is linear and non-Gaussian, then the bicoherence is a nonzero constant. Following binary hypothesis testing can be used for non-Gaussianity and nonlinearity detection:

$H_1$: the bispectrum of $y[n]$ is nonzero and not constant;

$H_0$: the bispectrum of $y[n]$ is nonzero and constant.

## V. EXPERIMENTAL RESULTS

### A. Dataset

Baidu Silicon Valley AI Lab cloned audio dataset is used for performance evaluation. It was downloaded from https://audiodemos.github.io. This dataset consists of 10 ground truth audio samples, 120 cloned recordings, and four morphed recordings. Summary of the dataset used is provided in Table I. More details about the dataset can be found in [15].

TABLE I. SUMMARY OF DATASET USED

| # OF SAMPLES | $VCTK_t \rightarrow VCTK_g$ | | $LibriSpeech_t \rightarrow VCTK_g$ | |
|---|---|---|---|---|
| | SEA | WMA | SEA | WMA |
| 1 | 4 | 4 | 4 | 4 |
| 5 | 4 | 4 | 4 | 4 |
| 10 | 4 | 4 | 4 | 4 |
| 20 | 4 | 4 | 4 | 4 |
| 50 | 4 | 4 | 4 | 4 |
| 100 | 4 | 4 | 4 | 4 |

| | Speaker encoder ($LibriSpeech_t \rightarrow VCTK_g$) | |
|---|---|---|
| | Without fine tuning | With fine tuning |
| 1 | 4 | 4 |
| 5 | 4 | 4 |
| 10 | 4 | 4 |

*SEA – Speaker Embedding Adaption*
*WMA – Whole Model Adaption*
*$LibriSpeech_t$ – training on Libri speakers*
*$VCTK_t$ – training on VCTK speakers*
*$VCTK_g$ – using VCTK speakers for cloned speech generation*

For this dataset generation, the multi-speaker model and speaker encoder model were trained on 84 VCTK speakers (48 KHz sampling rate), other VCTK speakers (48 KHz sampling rate) were used for voice cloning. The multi-speaker model and speaker encoder model were trained on LibriSpeech speakers (16 KHz sampling rate), VCTK speakers (down-sampled to 16 KHz samples/sec.) were used for voice cloning. The average duration of a cloning sample is 3.7 seconds.

### B. Results

Performance of the proposed fake audio detection method is evaluated using following set of experiments.

*1) Experiment 1:* The goal of this experiment is to evaluate performance of the proposed QPC-based detection on cloned audios generated through speaker embedding adaptation-based cloned audio generation discussed in [15]. The performance of Gaussianity and linearity test statistics based methods is also evaluated. To this end, 24x2 = 48 cloned recordings with whole model adaption along with 6 ground truth recordings downloaded from [15] are analyzed using HOSA. Both QPC and Gaussianity and linearity test statistics are estimated from cloned and ground truth recordings. Detection performance for both schemes is provided in the Table II. It can be observed from Table II that the poposed method successfully detected cloned audios.

*2) Experiment 2:* The goal of this experiment is to evaluate performance of the proposed QPC-based detection on cloned audios generated through whole model adaptation-based cloned audio generation discussed in [15]. In addition, performance of Gaussianity and linearity test statistics based methods is also evaluated for whole model adaptation-based cloned audio generation. To this end, 24x2 = 48 cloned recordings with whole model adaption along with 6 ground truth recordings downloaded from [15] are analyzed using HOSA. Both QPC and Gaussianity and linearity test statistics are estimated from cloned and ground truth recordings. Detection performance for both schemes is provided in the Table II. It can be observed from Table II that the poposed method successfully detected cloned recordings.

TABLE II. DETECTION PERFORMANCE

| | Detection Rate (%) | |
|---|---|---|
| | $VCTK_t \rightarrow VCTK_g$ | $LibriSpeech_t \rightarrow VCTK_g$ |

| # of samples | SEA | WMA | SEA | WMA |
|---|---|---|---|---|
| 1 | 100 | 100 | 100 | 100 |
| 5 | 100 | 100 | 100 | 100 |
| 10 | 100 | 100 | 100 | 100 |
| 20 | 100 | 100 | 100 | 100 |
| 50 | 100 | 100 | 100 | 100 |
| 100 | 100 | 75 | 100 | 75 |

*3) Experiment 3*: The goal of this experiment is to evaluate performance of the proposed method on cloned audios generated speaker encoding – based cloned speech synthesis as discussed in [15]. To this end, 12x2 = 24 cloned recordings along with three ground truth recordings. To this end, cloned speech recordings and three ground truth recordings download loaded are analyzed for QPC and Gaussianity and linearity test statistic estimation. The proposed methods successfully detected all 24 cloned recording and all three ground truth recordings.

*4) Experiment 4*: The goal of this experiment is to evaluate performance of the proposed method on morphed audios. The morphed audios are generated by manipulating estimated speaker embedding parameter by the speaker encoder as discussed in [15]. To this end, four morphed speech recordings along with two ground truth recordings download loaded are analyzed for QPC and Gaussianity and linearity test statistic estimation. The proposed methods successfully detected all four morphed recordings.

## VI. CONCLUSIONS

This paper presents a framework for cloned speech detection. We have demonstrated that cloned speech exhibit characteristic artifacts which are used for cloned audio detection. We have also demonstrated that cloned audios lacks higher order correlations. Higher-order spectral analysis is used for cloned audio detection. Specifically, QPC and Gaussianity and linearity test statistics are used for higher order correlation detection. Effectiveness of the proposed method is evaluated on cloned audios generated using speaker adaptation- and speaker encoding-based approaches proposed in [15]. Performance of the proposed for a dataset of 134 speech samples indicate that the proposed method is capable of detecting bona-fide and cloned audios with a perfect detection rate.

Our future work aims to rely on and traditional spectral features (spectrogram and MFCC) along with the higher order spectral features analysis to learn the underlying model for cloned audio and used it for cloned audio detection.

ACKNOWLEDGMENT

This research is supported by the National Science Foundation (NSF) under the awards No. 1816019.